\documentclass[9pt, conference]{IEEEtran}
\IEEEoverridecommandlockouts

\usepackage{cite}
\usepackage{amsmath,amssymb,amsfonts}
\usepackage{algorithmic}
\usepackage{graphicx}
\usepackage{textcomp}
\usepackage{xcolor}
\usepackage{booktabs, tabularx, tabularray, multirow}
\def\BibTeX{{\rm B\kern-.05em{\sc i\kern-.025em b}\kern-.08em
    T\kern-.1667em\lower.7ex\hbox{E}\kern-.125emX}}
\begin{document}

\title{EmotionCaps: Enhancing Audio Captioning Through Emotion-Augmented Data Generation
\thanks{This work was partially supported by the New Jersey Institute of Technology Honors Summer Research Institute (HSRI).}
}

\author{\IEEEauthorblockN{Mithun Manivannan}
\IEEEauthorblockA{\textit{Sound Interaction and Computing Lab} \\
\textit{New Jersey Institute of Technology}\\
Newark, USA \\
mm2356@njit.edu}
\and
\IEEEauthorblockN{Vignesh Nethrapalli}
\IEEEauthorblockA{\textit{Sound Interaction and Computing Lab} \\
\textit{New Jersey Institute of Technology}\\
Newark, USA \\
vbn2@njit.edu}
\and
\IEEEauthorblockN{Mark Cartwright}
\IEEEauthorblockA{\textit{Sound Interaction and Computing Lab} \\
\textit{New Jersey Institute of Technology}\\
Newark, USA \\
mark.cartwright@njit.edu}
}

\maketitle

\begin{abstract}

Recent progress in audio-language modeling, such as automated audio captioning, has benefited from training on synthetic data generated with the aid of large-language models. However, such approaches for environmental sound captioning have primarily focused on audio event tags and have not explored leveraging emotional information that may be present in recordings. In this work, we explore the benefit of generating emotion-augmented synthetic audio caption data by instructing ChatGPT with additional acoustic information in the form of estimated soundscape emotion. To do so, we introduce EmotionCaps, an audio captioning dataset comprised of approximately 120,000 audio clips with paired synthetic descriptions enriched with soundscape emotion recognition (SER) information. We hypothesize that this additional information will result in higher-quality captions that match the emotional tone of the audio recording, which will, in turn, improve the performance of captioning models trained with this data. We test this hypothesis through both objective and subjective evaluation, comparing models trained with the EmotionCaps dataset to multiple baseline models. Our findings challenge current approaches to captioning and suggest new directions for developing and assessing captioning models.
\end{abstract}


\section{Introduction}
\label{sec:intro}

Audio-language modeling tasks, such as language-based audio retrieval \cite{elizaldeCLAPLearningAudio2023,wuLargescaleContrastiveLanguageAudio2022,mei2023wavcaps}, language-guided source separation \cite{liuSeparateAnythingYou2023}, text-to-audio generation \cite{kongImprovingTextAudioModels2024a}, audio question answering (AQA) \cite{gongListenThinkUnderstand2023a,kongAudioFlamingoNovel2024,gardnerLLarkMultimodalFoundation2023}, and automated audio captioning (AAC) \cite{xuStatusQuoContemporary2024,meiAutomatedAudioCaptioning2022}, have received significant attention in recent years. However, while several audio-language datasets have been developed \cite{drossosClothoAudioCaptioning2020,kimAudioCapsGeneratingCaptions2019,martinmoratoDiversityBiasAudio2021}, access to an adequate amount of high-quality audio-language pairs is a persistent challenge for this area of research. To address this data scarcity problem and improve performance, researchers have proposed a number of text generation and augmentation schemes to increase the number of audio-caption pairs for training. For example, in two of the earliest approaches, Wu et al \cite{wuLargescaleContrastiveLanguageAudio2022} used a large-language model (LLM) to transform audio event tags to captions, and Mei et al \cite{mei2023wavcaps} further advanced this approach by also transforming noisy descriptions into typical audio captions. Alternative approaches have explored incorporating additional information into the data generation pipeline \cite{gongListenThinkUnderstand2023a,gardnerLLarkMultimodalFoundation2023,dohLPMusicCapsLLMBasedPseudo2023a,kongAudioFlamingoNovel2024}, retrieving additional captions for augmentation \cite{ghoshRecapRetrievalAugmentedAudio2024a}, and mixing-up existing captions for augmentation \cite{wuImprovingAudioCaptioning2023}. 

However, to our knowledge, none of these approaches have explored incorporating the ``emotion'' or ``mood'' of environmental sound as additional information in their data generation pipelines. Prior research has shown that soundscape emotion descriptors, e.g., \textit{pleasantness} and \textit{eventfulness}, are key components of soundscape perception \cite{axelssonPrincipalComponentsModel2010}, which have subsequently been linked to individuals' psychological well-being \cite{erfanianPsychologicalWellbeingDemographic2021}. Given their importance in soundscape perception and that many of the environmental sound recordings we may want to be captioned could be considered soundscapes, we hypothesize that the inclusion of this soundscape ``emotion'' information may improve the quality of synthetically generated captions. For example, without additional information, an LLM prompted given the AudioSet \cite{gemmekeAudioSetOntology2017} tag \textit{Vehicle horn, car horn, honking}, could generate a caption like ``A car horn honks in the distance.'' However, with additional information that indicates that the recording is \textit{highly chaotic}, an LLM may generate a caption like ``A car horn honks frantically amidst a chaotic scene'' --- a very different description of the auditory scene. 

In this paper, we explore our hypothesis by developing and evaluating a pipeline for emotion-augmented synthetic caption generation. First, we train a model to predict the soundscape emotion and magnitude. Next, we construct the EmotionCaps dataset by instructing ChatGPT \cite{openai2023gpt35turbo} to generate audio captions using audio event tags augmented with the predicted soundscape emotion information. Lastly, using EmotionCaps in combination with Clotho \cite{drossosClothoAudioCaptioning2020} and AudioCaps \cite{kimAudioCapsGeneratingCaptions2019}, we train a set of audio captioning models and compare their outputs both objectively and subjectively to multiple baseline models. Our findings challenge current approaches to captioning and suggest new directions for developing and assessing captioning models.

\section{EmotionCaps Dataset}
\label{sec:dataset}


\begin{figure}[t]
  \centering
  \centerline{\includegraphics[width=\columnwidth]{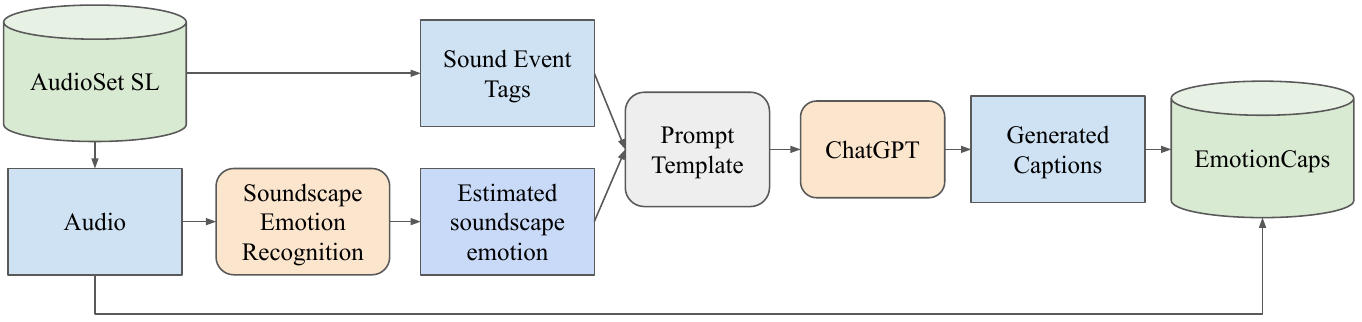}}
  \caption{EmotionCaps emotion-augmented caption generation pipeline}
  \label{fig:mocaflow}
\end{figure}

To incorporate soundscape emotion into audio captioning models, we developed the EmotionCaps dataset, which aims to bridge the gap between soundscape emotion recognition (SER) and automated audio captioning (AAC), using the following emotion-augmented audio caption generation pipeline: 

\subsection*{Step 1: Train a soundscape emotion recognition model}
We trained a model to predict the perceived emotion of audio clips using the Emo-Soundscapes dataset \cite{fanEmosoundscapesDatasetSoundscape2017}, which contains contains 1213 6-second audio clips sourced from Freesound \cite{akkermansFreesoundImprovedPlatform2011}---100 clips from each category in Schafer's taxonomy \cite{schaferSoundscapeOurSonic1993}:  \textit{natural sounds}, \textit{human sounds}, \textit{sounds and society}, \textit{mechanical sounds}, \textit{quiet and silence}, and \textit{sounds as indicators}; and 613 mixtures of those 600 clips. Each clip is labeled according to Russel's circumplex of affect \cite{russellCircumplexModelAffect1980}, i.e., valence and arousal, each on a scale from -1 to 1. Valence can be roughly thought of as a measurement of perceived pleasantness, while arousal is perceived eventfulness or energy. 
While the original model published in \cite{fanEmosoundscapesDatasetSoundscape2017} is not publicly available, we followed a similar modeling approach as they reported and trained a support vector regression (SVR) model on the principal components of 72 summary statistics (mean and standard deviation 13 MFCCs, 13 delta MFCCs, zero crossing rate, spectral centroid, spectral bandwidth, spectral contrast, spectral flatness, spectral roll off, loudness, delta loudness, root-mean-square energy, and delta root-mean-square energy). 
A grid search with cross-validation was used to choose the number of components for PCA, as well as to choose the C, gamma, and kernel parameters for the SVR. One such model was trained in the same way independently for valence and arousal. 

We evaluated the models similarly to those of Fan et al. \cite{Emo-Soundscapes}. In each trial, the dataset is shuffled and split into an 80:20 train-test split, and the model is trained and evaluated on $R^2$ and mean squared error (MSE). The aggregate results from 100 trials are shown in Table 1.

\begin{table}[t]
    \centering
    \caption{\normalfont{Fit of support vector regression model for estimating affect.}}
    \SetTblrInner{rowsep=0pt}
    \begin{tblr}{c|c|c|c|c}
        \hline
        \SetCell[r=2]{m}{} Affect Dimension & \SetCell[c=2]{c} $R^2$ && \SetCell[c=2]{c} $MSE$ \\ \hline
        
         & \SetCell{c} Mean & \SetCell{c} SD & \SetCell{c} Mean & \SetCell{c} SD \\ \hline
         
        Valence & 0.709 & 0.031 & 0.096 & 0.008 \\ \hline
        Arousal & 0.897 & 0.011 & 0.034 & 0.003 \\ \hline
    \end{tblr}
\end{table}

\begin{table*}[t]
    \caption{\normalfont{Prompt variations used to generate captions in EmotionCaps.}}
    \centering
    \begin{tblr}{c|p{0.7\textwidth}}
        \hline
        Prompt Method & \SetCell{c}{} Prompt \\ \hline
        \SetCell[r=1]{m} WavCaps & \SetCell[r=1]{m} \textit{I will give you a number of lists containing sound events occurred sequentially in time. Process each individually. Write a one-sentence audio caption to describe these sounds. Make sure you are using grammatical subject-verb-object sentences. Directly describe the sounds and avoid using the word "heard". The caption should be less than 20 words.} \\ \hline
        
        \SetCell[r=1]{m} Scene-focused & \SetCell[r=1]{m} \textit{I will provide a list containing chronological sound events of an auditory scene. Write a one-sentence audio caption to describe the scene. Make sure to use an active voice. Describe the scene without simply listing the sounds. The caption should be less than 20 words.} \\ \hline
        
        Emotion Addon & (Appended to Scene-focused) \textit{I will also provide a mood. Please emphasize this mood in your caption.} \\ \hline
        
        \SetCell[r=1]{m} Emotion Rewrite & \SetCell[r=1]{m} (Following to Scene-focused response) \textit{I will give you a sentence describing a sound scene, and a mood. Please rewrite the sentence, emphasizing the indicated mood.}
        \label{tab:prompts}
   \end{tblr}

  \vspace{4mm}
  \caption{\normalfont{Example captions given sound event tags and estimated emotion for different prompt variations in EmotionCaps.}}
  \begin{tblr}{c|p{0.34\textwidth}|p{0.34\textwidth}}
      \hline
      \SetCell[r=2]{m} Prompt Method & \SetCell[c=2]{c} Sound Events and Estimated Emotion \\ \hline

       & \SetCell{c} {['Thunder', 'Rain on surface'] \\ Emotion: unpleasant} & \SetCell{c} [`Wind', `Rustle', `Bird vocalization, bird call, bird song']; \ Emotion: uneventful \\ \hline

       \SetCell[r=1]{m} WavCaps & \SetCell[r=1]{m} Thunder rumbles as rain falls on the surface. & \SetCell[r=1]{m} The wind blows, causing rustling sounds, while birds vocalize with their calls and songs. \\ \hline

       \SetCell[r=1]{m} Scene-focused & \SetCell[r=1]{m} Thunder rumbles as rain falls on a surface. & \SetCell[r=1]{m} The wind rustles as birds sing their melodic calls. \\ \hline

       \SetCell[r=1]{m} Emotion Addon & \SetCell[r=1]{m} Thunder rumbles ominously as rain pounds relentlessly on the surface. & \SetCell[r=1]{m} The wind gently rustles as birds sing, creating a peaceful and uneventful atmosphere. \\ \hline

       \SetCell[r=1]{m} Emotion Rewrite & \SetCell[r=1]{m} Thunder ominously rumbles as relentless rain falls on a desolate surface. & \SetCell[r=1]{m} The wind softly rustles as birds peacefully sing their melodic calls. 
       \label{tab:example_chatgpt_outputs}
  \end{tblr}
\end{table*}

To express the emotion with language, we discretized each of the two-dimensional output vectors into one of 8 emotions (\textit{eventful}, \textit{uneventful}, \textit{pleasant}, \textit{unpleasant}, \textit{exciting}, \textit{boring}, \textit{quiet}, and \textit{chaotic}---emotions equally spaced around the unit circle) by finding its nearest neighbor via cosine similarity. Furthermore, we projected the output vector onto its nearest neighbor emotion vector and retained the magnitude of the projection as an indicator of the emotion intensity. We discretized these intensities into emotion qualifiers based on percentile scores, $p$, in the sample distribution: $p < 0.15$: ``neutral'', $0.15 <= p < 0.50$: ``slightly \textit{\textless emotion\textgreater}'', $0.5 <= p < 0.85$: ``\textit{\textless emotion\textgreater}'', $0.85 <= p <= 1.0$: ``highly \textit{\textless emotion\textgreater}''.

\subsection*{Step 2: Augment sound event annotations with estimated soundscape emotion}
While audio events could be estimated with a SED model, we opted to eliminate that potential source of error in our experiments by instead leveraging the ground-truth annotations in AudioSet SL~\cite{hersheyBenefitTemporallyStrongLabels2021}, the strongly-labeled set of 120,071 audio clips from the larger AudioSet \cite{gemmekeAudioSetOntology2017} dataset. We converted the sound event annotations into a temporally ordered list of sound events, retaining only the first chronological occurrence of each unique sound event in an audio clip. This sound event list was then augmented using the qualified emotion labels estimated for the model trained in Step 1. 


\subsection*{Step 3: Instruct an LLM to construct caption sentences given the estimated emotion along and sound event tags}
Mei et al. demonstrated that ChatGPT \cite{openai2023gpt35turbo} can be a valuable tool in generating synthetic captions for training audio-language models \cite{meiWavCapsChatGPTAssistedWeaklyLabelled2023}. We leverage that work by using the WavCaps prompt for AudioSet SL as a base prompt upon which we build with our predicted emotions. We create three new prompt variations. The first modified the WavCaps prompt to first describe the scene and only asked for one caption at a time. We refer to this prompt as \textit{Scene-focused}. Next, we appended that prompt with a statement that we would also include a mood with the list with the intention of generating the desired caption with one instruction. We call this \textit{Emotion Addon}. Lastly, we created a prompt variation in which we instructed the LLM in two steps, first with the scene-focused prompt and then with a prompt instructing to rewrite the first response with a given emotion. We refer to this prompt method as \textit{Emotion Rewrite}. We used these four prompts, along with the AudioSet SL sound events and our predicted emotions, to instruct ChatGPT-3.5 Turbo and create four dataset variations. Note that we chose to generate new captions with the WavCaps prompt rather than use examples from the published WavCaps dataset in order to eliminate variation due to changes in ChatGPT versions. See Table~\ref{tab:prompts} for the prompts and Table~\ref{tab:example_chatgpt_outputs} for example outputs from ChatGPT.

\subsection{Dataset Overview}
\label{sec:dataanalysis}

The resulting EmotionCaps dataset\footnote{https://doi.org/10.5281/zenodo.13755932} contains four subsets (one for each prompt method) of captions for the 120,071 audio clips from AudioSet SL. The average word counts for captions from the WavCaps prompt (12.61), scene-focused base prompt (14.04), emotion addon prompt (18.35), and emotion rewrite prompt (18.65). The difference between WavCaps and the emotion prompts demonstrates the difference in sentence length when infusing captions with emotion information.

\section{Experiments}
\label{sec:experiments}

\subsection{Datasets}
\label{sec:datasets}
We used three AAC datasets when training and evaluating the AAC models in our experiments: AudioCaps \cite{kimAudioCapsGeneratingCaptions2019}, Clotho \cite{drossosClothoAudioCaptioning2020}, and EmotionCaps. AudioCaps consists of approximately 51,308 10 s clips from AudioSet, each paired with 1 human-labeled caption for the training split and 5 for the test split. Clotho consists of 4981 15--30 s clips, each paired with 5 human-labeled captions in both training and test splits.

\subsection{Architecture}
\label{sec:models}
To evaluate the EmotionCaps dataset on the AAC task, we employed the same encoder-decoder architecture used to evaluate WavCaps, which was derived from the DCASE 2022 Challenge Task 6a baseline model \cite{dcase2022task6a} and consists of an HTSAT \cite{chenHTSATHierarchicalTokenSemantic2022} Trasformer-based audio-encoder pre-trained on AudioSet classification and a BART \cite{lewis-etal-2020-bart} Transformer-based language decoder pre-trained on large text corpora.

\subsection{Training}
\label{sec:training}
We utilized a two-stage training process mirroring that of WavCaps AAC training. The first stage consists of pretraining on EmotionCaps (only one subset for each model), combined with the Clotho and AudioCaps training sets. The first stage used a learning rate of $5\times 10^{-5}$ and a batch size of 48 for 15 epochs. The second stage fine-tuned the model further on either AudioCaps and Clotho when evaluated on the AudioCaps and Clotho test sets, respectively. The second stage used a learning rate of $5\times 10^{-6}$ with a batch size of 32 for 20 epochs. This fine-tuning is necessary for the objective evaluation since the style of the two test sets is a bit different---Clotho tends to exhibit more imaginative description, using more sophisticated vocabulary phrasing \cite{xuStatusQuoContemporary2024}. Since EmotionCaps features 4 different caption subsets, 4 separate models were trained such that each used a different caption subset. As another baseline, we also trained a fifth model without synthetic data, exclusively on both AudioCaps and Clotho for stage 1, and followed the same training as the other models in stage 2. From hereon, we refer to the model trained on the WavCaps prompted subset as ``WavCaps-Like" since this, of course, is a different model than that published in the WavCaps paper, which was trained on synthetic caption data generated from three additional datasets (including 262k clips from FreeSound~\cite{akkermansFreesoundImprovedPlatform2011}) which is significantly larger than AudioSet SL) than we incorporate into this work. A limitation of this current work is that synthetic data was only generated using AudioSet SL. This likely limits overall performance, but the comparisons between data generation strategies should not be affected by this. 

\subsection{Objective Evaluation Metrics}
\label{sec:objectivemetrics}
To evaluate the objective performance of the five models, we used the METEOR \cite{banerjee-lavie-2005-meteor}, CIDEr \cite{CIDEr}, SPICE \cite{anderson2016spicesemanticpropositionalimage}, SPIDEr \cite{SPIDEr}, 
and FENSE \cite{FENSE} metrics as implemented in \texttt{aac\_metrics}\cite{Labbe_aac_metrics_2024} to evaluate the tests of Clotho and AudioCaps. FENSE computed on the validation split was used for model selection during training. We evaluated the best checkpoints from both stage 1 and stage 2.

\subsection{Subjective Listening Test}
\label{sec:subjective_listening_test}
To evaluate the effectiveness of incorporating emotional data into captions, we also conducted a subjective listening test to determine whether participants preferred emotionally enriched captions over other variations. We selected 30 random examples from the AudioCaps test set, which were divided into six stimulus groups of five examples each, and we used the five models described in Section~\ref{sec:training} to estimate captions for each example. Note that in the subjective evaluation, we only used checkpoints after stage 1 training, i.e., we didn't fine-tune on AudioCaps or Clotho---this choice was made so as to retain the emotion information from stage 1 and not fit to the style of one particular dataset. In addition, for each audio example, we added two randomly selected ground-truth captions (of the five) to the set, as well as one random caption from the AudioCaps test set that was sufficiently different from the ground-truth. This latter was done by computing the cosine similarity between the sentence embeddings\cite{song2020mpnet}\footnote{https://huggingface.co/sentence-transformers/all-mpnet-base-v2} of ground-truth captions and all the other captions in the test set and then randomly selecting one for which the similarity averaged over both ground-truth captions was less than the median similarity of the sample. We recruited a total of 30 participants via the Prolific crowd-sourcing platform, all of whom were native English speakers from the United States with no reported known hearing conditions. Participants were randomly assigned to one of the 6 stimulus groups We asked participants to listen to an audio sample and then rank order the eight captions on four criteria from best to worst: \textit{preference}, \textit{accuracy}, \textit{completeness}, and \textit{affect}. Descriptions and examples were used to clarify the scales. The \textit{accuracy} scale was the precision metric described in layman's terms; The \textit{completeness} scale was the recall metric described in layman's terms; and the \textit{affect} scale was described as ``how well the caption conveys the emotional tone of the audio.'' This was task repeated for each audio example in a group in random order. In total, each caption was ranked a total of 5 times for each scale. Participants were compensated \$3 and the median completion time was roughly 15 minutes.

\begin{figure}[t]
  \centering
  \centerline{\includegraphics[width=\columnwidth]{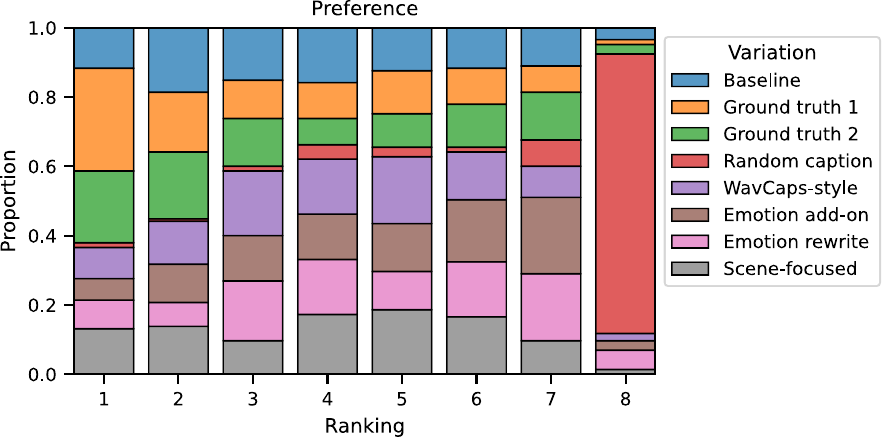}}
  \caption{Preference rank distribution in subjective listening test.}
  \label{fig:preference-distribution}
\end{figure}

\begin{table}[t]
\small
    \centering
    \caption{\normalfont{Mean Ranking of Captions On Subjective Scales. Bold indicates best AAC system per scale. Lower is better.}}
    \SetTblrInner{rowsep=0pt}
    \begin{tblr}{c|c c c c }
        \hline
        \textbf{Variation} & \textbf{Preference} & \textbf{Acc.} & \textbf{Completeness} & \textbf{Affect} \\
        \hline
        GT 1  & 3.27 & 3.37 & 3.66 & 4.05 \\
        GT 2  & 3.72 & 3.90 & 4.15 & 4.40 \\
        \hline
        Baseline       & \textbf{3.95} & \textbf{3.81} & 4.17 & 4.58 \\
        \hline
        WavCaps-Like  & 4.12 & 4.32 & 4.16 & 4.12 \\
        Scene-Focused  & 4.10 & 3.97 & \textbf{3.92} & 4.01 \\
        Emotion Addon     & 4.73 & 4.81 & 4.19 & 3.81 \\
        Emotion Rewrite   & 4.67 & 4.44 & 4.37 & \textbf{3.68} \\
        \hline
        Random  & 7.44 & 7.38 & 7.38 & 7.35 \\
        \hline
    \end{tblr}
    \label{tab:caption_ranking}
    \vspace{4mm}
    
    \caption{\normalfont{Objective AAC results on test sets of the AudioCaps (AC) and Clotho (Cl) datasets (DS). Higher is better on all metrics. Models: Scene-Focused (S-F), Emotion Addon (E-A), Emotion Rewrite (E-RW), WavCaps-Like (WC-L), Baseline (B).}}    
    \centering
    \SetTblrInner{rowsep=0pt,colsep=3pt}
    \begin{tblr}{ c | c *{7}{c}}
        \hline
         \textbf{DS} & \textbf{Stage} & \textbf{Var.} & \textbf{METEOR} & \textbf{CIDEr} & \textbf{SPICE} & \textbf{SPIDEr} & \textbf{FENSE} \\
        \hline
        \SetCell[r=10]{m} Cl & \SetCell[r=5]{m} 1 & S-F & 14.4 & 26.7 & 9.67 & 18.19 & 43.7 \\
         & & E-A & 13.5 & 21.6 & 9.0 & 15.3 & 42.3 \\ 
         & & E-RW & 13.7 & 24.7 & 9.4 & 17.0 & 44.5 \\
         & & WC-L & 13.8 & 25.0 & 8.9 & 17.0 & 44.3 \\
         & & B & \textbf{16.1} & \textbf{33.0} & \textbf{10.8} & \textbf{21.7} & \textbf{47.6} \\ \hline
         & \SetCell[r=5]{m} 2 & S-F & 17.3 & 38.8 & 12.5 & 25.7 & 46.0 \\
         & & E-A & 17.7 & 38.5 & 12.4 & 25.4 & 47.0 \\
         & & E-RW & 17.7 & \textbf{40.4} & \textbf{12.6} & \textbf{26.5} & \textbf{48.3} \\
         & & WC-L & 9.1 & 28.3 & 11.0 & 19.7 & 46.2 \\
         & & B & \textbf{18.0} & 36.8 & 12.5 & 24.6 & 48.0 \\ \hline
        \SetCell[r=10]{m} AC & \SetCell[r=5]{m} 1 & S-F & 20.2 & 33.1 & 14.7 & 23.9 & 57.2 \\
         & & E-A & 15.9 & 27.0 & 11.9 & 19.4 & 46.7 \\
         & & E-RW & 17.5 & 34.0 & 12.2 & 23.1 & 49.6 \\
         & & WC-L & 20.5 & 35.6 & 14.2 & 24.9 & 56.6 \\
         & & B & \textbf{23.5} & \textbf{70.1} & \textbf{17.8} & \textbf{44.0} & \textbf{60.8} \\ \hline
         & \SetCell[r=5]{m} 2 & S-F & 22.8 & 63.8 & 17.2 & 40.5 & 61.9 \\
         & & E-A & \textbf{24.2} & \textbf{73.8} & \textbf{17.8} & \textbf{45.8} & \textbf{62.2} \\
         & & E-RW & 24.1 & 71.2 & 17.4 & 44.3 & 61.7 \\
         & & WC-L & 22.5 & 59.9 & 15.9 & 37.9 & 60.8\\
         & & B & 23.9 & 73.1 & 17.7 & 45.4 & 61.4\\ \hline
    \end{tblr}
    \label{tab:objective_results}
\end{table}

\section{Results}
\label{sec:results}

\subsection{Subjective evaluation} As can be seen in Table~\ref{tab:caption_ranking}, the random caption consistently has the highest mean ranking (i.e., ranked last), and while the random caption was occasionally not ranked last (see Figure~\ref{fig:preference-distribution}), these occurrences are infrequent. These observations provide some validity to our listening test. 

We find that captions from models trained on emotion-enriched captions are ranked best on \textit{affect}. Thus, we achieved our goal of training an AAC model to match the emotional tone of soundscapes.  Small differences in prompting also seem to lead to consistently better results---prompting in two steps seems a better strategy than one step (mean affect ranking of Emotion Rewrite: 3.68 vs. Emotion Addon: 3.81), and the small changes made in the Scene-Focused from the WavCaps prompt also consistently improved the model on all scales.  

Unfortunately, the emotion-enriched captions performed the worst of the AAC models on \textit{completeness} and \textit{accuracy}---though, all models performed quite similarly on \textit{completeness}, which exhibited the smallest range of mean ranking of all the scales. The performance of emotion-enriched captions was also the lowest of the AAC models for \textit{preference}, but if we look at the distribution of rankings, we see that despite their overall preference, the ground-truth labels were not consistently ranked the best---53\% of the time an AAC model was preferred, and more specifically 12\% of the time an emotion-enriched model was preferred.

\subsection{Objective evaluation}
When we evaluate the captioning models on the objective metrics (see Table~\ref{tab:objective_results}), we find that the baseline model is consistently the highest forming model in stage 1 for both datasets, indicating that training on out-of-distribution synthetic data (as all the other models do) is not helpful without fine-tuning since it likely pushes the outputs further away from the distribution of the test data. With the fine-tuning of stage 2, we see that the baseline model's performance typically only minimally improves. However, the models trained with synthetic data in stage 1 often saw a dramatic increase in performance in stage 2, so much so that we see that the emotion-enriched models actually performed the best on average in stage 2, with the Emotion Addon model performing best on AudioCaps, and the Emotion Rewrite performing best on Clotho. However, the improvements over the baseline are minimal. Similar to the subjective evaluation results, we also observe that the Scene-Focused model exhibits slightly better performance than the WavCaps-Like model.

\section{Discussion}
The emotion-enriched caption models achieved their goal of communicating the emotional tone of the soundscape through the captions. However, their improvement over the baseline was minimal in the objective evaluation---this, however, is not surprising given that they likely don't match the descriptive style of those test sets. In general, they were not preferred over the baseline in the subjective evaluation, but the distribution of rankings in preference indicates that preference for captions varies and that emotion-enriched captions are preferred by some people. This observation indicates maybe we should re-think how we train and evaluate AAC models. Is the one-size-fits-all approach adopted by AAC research the right one? If research in accessible speech captioning \cite{arroyochavezCustomizationClosedCaptions2024} and image description \cite{stanglGoingOneSizeFitsAllImage2021} is indicative, then the answer is ``no''---users have varied preferences and needs for how media should be captioned, and thus customizable and adaptable approaches should possibly be adopted in AAC as well. Future work should investigate these needs and inform how we should design, train, and evaluate AAC models.

\section{Conclusion}
\label{sec:conclusion}
We proposed a emotion-augmented data generation pipeline for training emotion-enriched AAC models. Emotions are estimated with a soundscape emotion recognition model, and ChatGPT is instructed to generate captions given the estimated emotions and sound event tags. Our experiments found that the emotion-enriched AAC models generate captions that match the emotional tone of the input audio more than the baseline models we compared, and the emotion-enriched models performed slightly better than baseline models on standard AAC captioning metrics. The emotion-enriched models were not generally preferred in our subjective listening tests, but our study results exhibit that people have varied preferences on captions. We propose that future work should investigate the diversity of captioning needs and revisit how we design, train, and evaluate AAC models.

\bibliographystyle{IEEEtran}
\bibliography{IEEEabrv,refs}

\end{document}